\documentclass[twocolumn]{aastex63}

\newcommand{\chg}[1]{{\color{black}       #1}}

\received{February 7, 2022}
\revised{May 6, 2022}
\accepted{May 7, 2022}
\shorttitle{The Heartworm Nebula}
\shortauthors{Cotton et al.}

\begin{document}

\title{The Curious Case of the ``Heartworm'' Nebula}

\correspondingauthor{William Cotton}
\email{bcotton@nrao.edu}

\author{W.~D.~Cotton}
\affiliation{National Radio Astronomy Observatory,
	520 Edgemont Road,
Charlottesville, VA 22903, USA}
\affiliation{South African Radio Astronomy Observatory,
	2 Fir Street,
Observatory 7925, South Africa}

\author{F.~Camilo}
\affiliation{South African Radio Astronomy Observatory,
	2 Fir Street,
Observatory 7925, South Africa}

\author{W.~Becker}
\affiliation{Max-Planck-Institut f\"ur extraterrestrische Physik,
Giessenbachstra{\ss}e 1, 85748 Garching, Germany}
\affiliation{Max-Planck-Institut f\"ur Radioastronomie,
Auf dem H\"ugel 69, 53121 Bonn, Germany}

\author{J.~J.~Condon}
\affiliation{National Radio Astronomy Observatory,
	520 Edgemont Road,
Charlottesville, VA 22903, USA}

\author{J.~Forbrich}
\affiliation{Centre for Astrophysics Research, University of
Hertfordshire, College Lane, Hatfield AL10 9AB, UK}

\author{I.~Heywood}
\affiliation{Astrophysics, Department of Physics, University of
Oxford, Keble Road, Oxford OX1 3RH, UK}
\affiliation{South African Radio Astronomy Observatory,
	2 Fir Street,
Observatory 7925, South Africa}
\affiliation{Department of Physics and Electronics, Rhodes University,
PO Box 94, Makhanda 6140, South Africa}

\author{B.~Hugo}
\affiliation{South African Radio Astronomy Observatory,
	2 Fir Street,
Observatory 7925, South Africa}

\author{S.~Legodi}
\affiliation{South African Radio Astronomy Observatory,
	2 Fir Street,
Observatory 7925, South Africa}

\author{T.~Mauch}
\affiliation{South African Radio Astronomy Observatory,
	2 Fir Street,
Observatory 7925, South Africa}

\author{P.~Predehl}
\affiliation{Max-Planck-Institut f\"ur extraterrestrische Physik,
Giessenbachstra{\ss}e 1, 85748 Garching, Germany}

\author{P.~Slane}
\affiliation{Center for Astrophysics, Harvard \& Smithsonian,
60 Garden Street, Cambridge, Massachusetts 02138, USA}

\author{M.~A.~Thompson}
\affiliation{School of Physics and Astronomy, University of Leeds,
Leeds LS2 9JT, UK}



\begin{abstract}

The curious Galactic features near G357.2$-$0.2 were observed with
the MeerKAT radio interferometer array in the UHF and L bands
(0.56--1.68 GHz).  There are two possibly related features: a newly
identified faint heart-shaped partial shell (the ``Heart''), and a
series of previously known but now much better imaged narrow, curved
features (the ``Worm'') interior to the heart.  
Polarized emission suggests that much of the emission is nonthermal
and is embedded in a dense plasma.  The filaments of the worm appear
to be magnetic structures powered by embedded knots that are sites
of particle acceleration.  The morphology of the worm broadly
resembles some known pulsar wind nebulae (PWNe) but there is no
known pulsar or PWN which could be powering this structure. We also
present eROSITA observations of the field; no part of the nebula
is detected in X-rays, but the current limits do not preclude the
existence of a pulsar/PWN of intermediate spin-down luminosity.

\end{abstract}



\section{Introduction} \label{sec:intro}

\cite{Broadbent89} identified a feature near G357.2$-$0.2 (G357.1$-$00.2
in some references) as a candidate supernova remnant (SNR) because
its $S_{\rm 60\,\mu m}/S_{\rm 6\,cm}$ flux-density ratio is lower
than that of Galactic H\textsc{ii} regions and it is resolved at
6~cm with the Parkes telescope 4\arcmin\ beam.  \cite{Gray94} added
the 1\arcmin\ resolution 843\,MHz Molonglo Observatory Synthesis
Telescope image clearly resolving a sinuous structure for the first
time and indicating a nonthermal radio spectrum.  There is a diffuse
halo surrounding the fine scale structure.

\cite{Gray96} was the first to present and discuss high-resolution
(13\arcsec) Very Large Array images of this nebula.  The author
also noted the high polarization of the filaments at C band (5\,GHz)
and the low polarization at L band (1.5\,GHz), indicating depolarization
and rotation measure $\mbox{RM}\sim2000$\,rad\,m$^{-2}$.  On the
basis of the unusual morphology, \cite{Gray96} deprecated the SNR
interpretation and mentioned a variety of possibilities, including
a pulsar wind nebula (PWN) and one more example of peculiar nonthermal
phenomena near the Galactic center \citep[e.g., the ``Tornado''
only $0\fdg5$ away from G357.2$-$0.2,][]{Gaensler2003}.

\cite{Gray94} and \cite{Gray96} note that the pulsar B1736$-$31 is
in the vicinity of G357.2$-$0.2, in projection.  Its location outside
the nebula precludes any connection to a PWN interpretation, and
its spin-down age of 0.5\,Myr \citep{Clifton92} also makes it too
old to still have an associated visible SNR.

H\textsc{i} observations of G357.2$-$0.2 by \cite{Roy2002} give a
distance of at least 6\,kpc and place it either in front of, or
partly embedded in, a cloud believed to be beyond the Galactic
center; they conclude that it is Galactic.

We observed G357.2$-$0.2 with the MeerKAT radio telescope\footnote{Operated
by the South African Radio Astronomy Observatory (SARAO).} in the
UHF and L bands (0.56--1.68\,GHz) with 7\arcsec\ resolution and
with the eROSITA X-ray instrument.  The observations and analysis
are described in Section \ref{ObsAnalysis}, the imaging results are
presented in Section \ref{Results}, and a discussion of these results
is in Section \ref{Discussion} followed by a summary in Section
\ref{Summary}.

\section{Observations and Data Analysis \label{ObsAnalysis}}

\subsection{MeerKAT Observations, Analysis, and Imaging}

We observed G357.2$-$0.2 in both ``L'' (886--1682\,MHz) and UHF
(563--1068\,MHz) bands with the 64 antenna MeerKAT array pointed
at J2000 $\mbox{R.A.} = 17^{\mathrm h}39^{\mathrm m}39\fs82$,
$\mbox{Dec.} = -31\arcdeg27\arcmin47\farcs0$ (G357.176$-$0.235).
The integration time was 8\,s, and each band was divided into 4096
spectral channels.

The observations were in two sessions, L band on 2020 July 21 for
8 hours with 59 antennas and UHF on 2020 August 18 for 8 hours with
53 antennas.  PKS~B1934$-$638 was used as the flux density, band-pass
and delay calibrator, 3C~286 as the polarization calibrator, and
J1830$-$3602 as the astrometric calibrator.  The observing sequence
cycled between J1830$-$3602 (2 minutes) and G357.2$-$0.2 (20 minutes)
with a flux/band-pass calibrator (10 minutes) every 2 hours.  Our
flux-density scale is based on the \cite{Reynolds94} spectrum of
PKS~B1934$-$638: \begin{equation}
  \log(S) = -30.7667 + 26.4908 \bigl(\log\nu\bigr) - 7.0977
  \bigl(\log\nu\bigr)^2 $$ $$+0.605334 \bigl(\log\nu\bigr)^3,
\label{eq:pks1934} \end{equation} where $S$ is the flux density in
Jy and $\nu$ is the frequency in MHz.

\subsubsection{Analysis}

Data flagging and calibration were performed as described for L-band
data in \cite{DEEP2} and \cite{XGalaxy}.  The UHF session was
calibrated independently, and we have adopted the L-band procedure
for the UHF data with some band-specific modifications described
below.

First, we trimmed 144 channels from each edge of the UHF band to
account for the roll-off in receiver response, leaving a frequency
range 563--1069\,MHz. We then used a UHF-specific mask to identify
frequency ranges that contain persistent and strong radio frequency
interference (RFI). This covers only 934--960\,MHz, where cellular
communication signals are present.  After combining our empirical
mask with the editing steps described in \cite{DEEP2} during
calibration, $\sim 10\%$ of the target data were flagged from the
trimmed UHF band.

The data were split into 8 sub-bands with equal frequency width and
these were calibrated independently. We used a UHF sky model
extrapolated from the L-band model of the PKS~B1934$-$638 field
containing the power-law spectra of sources appearing brighter than
1\,mJy\,beam$^{-1}$ at 1.3\,GHz within $1^\circ$ of PKS~B1934$-$638.
The flux density of PKS~B1934$-$638 in each sub-band was obtained
from equation~(1), and used to derive the amplitude spectrum of
J1830$-$3602. The amplitudes of the gains measured from J1830$-$3602
were scaled by a smooth model fitted to its measured flux densities
in each sub-band, and the scaled amplitude and phase corrections
were interpolated in time and applied to the target data. The data
were reweighted using the root mean square (RMS) in the observed
visibilities in 10 minute intervals.

The above extrapolation does not account for sources towards the
edge of the wider UHF field of view (FoV). However, we have compared
the above analysis to one that uses a preliminary model of the full
UHF FoV of PKS~B1934$-$638, and find no appreciable difference in
the derived flux scales above 700\,MHz. Below this frequency our
derived flux densities are somewhat (up to 10--20\%) overestimated.

Imaging used the wide-band, wide-field imager MFImage in the
\emph{Obit}
package\footnote{\url{http://www.cv.nrao.edu/~bcotton/Obit.html}}
\citep{Obit} as described in \cite{DEEP2} and \cite{XGalaxy}.
MFImage \citep[described in detail in][]{SourceSize} uses faceting
to account for the non-coplanarity of the MeerKAT baselines and
multiple frequency bins which are imaged independently and CLEANed
jointly to account for frequency variations in the sky and the
antenna pattern.  Imaging used Robust weighting ($-1.5$ in
\emph{AIPS}/\emph{Obit} usage) to down-weight the central condensation of
antennas in the array and improve the resolution.

\subsubsection{Total-Intensity Imaging}\label{StokesIimaging}

The data in the two frequency bands were imaged independently.  With
the large bandwidth covered by the data, the shortest baseline
length in wavelengths varied by a factor of three between the highest
and lowest frequencies in the two bands.  Due to the large-scale
emission in the field, if uncorrected, this will lead to a variable
fraction of the total intensity recovered as a fraction of frequency
and a frequency-dependent negative bowl around the extended emission.
This will artificially cause the spectrum to appear steeper than
it actually is.  In order to counteract this, an inverted Gaussian
taper centered at the origin was applied to the weights of the
shortest baselines with a Gaussian $\sigma$ of 500 wavelengths to
both the UHF and the L-band data.  
This will suppress emission on scales larger than $\sim$200\arcsec;
this is similar to the spectral index analysis in \cite{XGalaxy}.  
A multi-resolution CLEAN was used to help recover the very extended
emission in the field.   

The L-band total-intensity data were imaged to a radius of 1$^\circ$
plus outlier facets to a distance of $1\fdg5$ to cover sources
expected to appear brighter than 1\,mJy\,beam$^{-1}$ based on the
SUMSS catalog at 843\,MHz \citep{SUMSS}.  Three iterations of
phase-only self-calibration were applied.  The total band-pass was
divided into 14 $\times$ 5\% fractional bandwidth bands giving
unequal widths in frequency.  L-band total-intensity imaging used
366,886 components stopping at a depth of 45\,$\mu$Jy\,beam$^{-1}$
with a total flux density of 23.7\,Jy; the off-source RMS noise is
20\,$\mu$Jy\,beam$^{-1}$.  The CLEAN restoring beam was an elliptical
Gaussian with FWHM axes $7\farcs0 \times 6\farcs8$ at position angle
0$^\circ$.

At UHF a field of view with radius $2\fdg5$ was imaged in 14 $\times$ 5\%
fractional bands with phase self-calibration using 419,484 components
to a minimum of 200\,$\mu$Jy\,beam$^{-1}$ and a total flux density
of 60.9\,Jy.  Outliers were added up to $3\fdg5$ from the pointing.
The off-source RMS was 89\,$\mu$Jy\,beam$^{-1}$.  The CLEAN restoring
beam was $11\farcs6 \times 10\farcs3$ at position angle $-20^\circ$.

For both L band and UHF, the 8\,s integrations and sub-bands used
introduce negligible time and bandwidth smearing ($<2\arcsec$)
across the full imaged FoVs.

\subsubsection{Deconvolution of Stokes Q and U}

Only the L-band data had adequate polarization calibration and were
imaged in Stokes Q and U.  In order to recover the polarimetry in
the presence of the large Faraday rotation of polarized emission,
a relatively high spectral resolution was used for Stokes Q and U
imaging --- a 1\% fractional bandwidth resulting in 68 sub-bands
across the band.  The deconvolution also used the joint polarization
CLEAN described in \cite{Condon2021}.  Linear polarization imaging
used 50,000 CLEAN components to a depth of 54\,$\mu$Jy\,beam$^{-1}$
resulting in an off-source RMS of 10\,$\mu$Jy\,beam$^{-1}$.

\subsection{eROSITA Observations and Analysis} \label{sec:eROSITA}

The X-ray eROSITA \citep[extended R\"ontgen Survey Imaging Telescope
Array,][]{Predehl2020a} is one of two instruments on the Spectrum
R\"ontgen-Gamma observatory \citep{Sunyaev2021}.  It consists of
seven aligned X-ray telescopes (TM1--TM7) which have an FoV of
1\degr.  All telescopes observe the same sky region simultaneously
in the 0.2--8\,keV band-pass.  In survey mode, the instrument's
angular resolution is $26\arcsec$. eROSITA started its first all-sky
survey on 2019 December 13, with eight such surveys planned over 4
years \citep[see][]{Predehl2020a}.

The X-ray data we report here were taken during the first four
eROSITA surveys, eRASS:4. By end 2021 the position of G357.2$-$0.2
had been observed with a total of 27 telescope passages during four
epochs, 2020 March 27--28, 2020 September 28--30, 2021 March 24--25,
and 2021 September 24--25, resulting in an un-vignetted averaged
exposure time of 1048\,s.

The data used in our analysis were processed by the eROSITA Standard
Analysis Software System (\emph{eSASS}) pipeline and have the
processing number $\#946$. For the data analysis we used \emph{eSASS}
version 201009\footnote{See \url{https://erosita.mpe.mpg.de/}}.
Within the \emph{eSASS} pipeline, X-ray data of the eRASS sky are
divided into 4700 partly overlapping sky tiles of $3\fdg6 \times
3\fdg6$ each. These are numbered using six digits, three each for
R.A. and Dec., encoding the sky tile center position in degrees.
The majority of G357.2$-$0.2 falls into the eRASS tiles 266120 and
266123, with the surrounding tile 263123 also required for a complete
coverage of G357.2$-$0.2.

\section{Results\label{Results}}

\begin{figure*}
\plotone{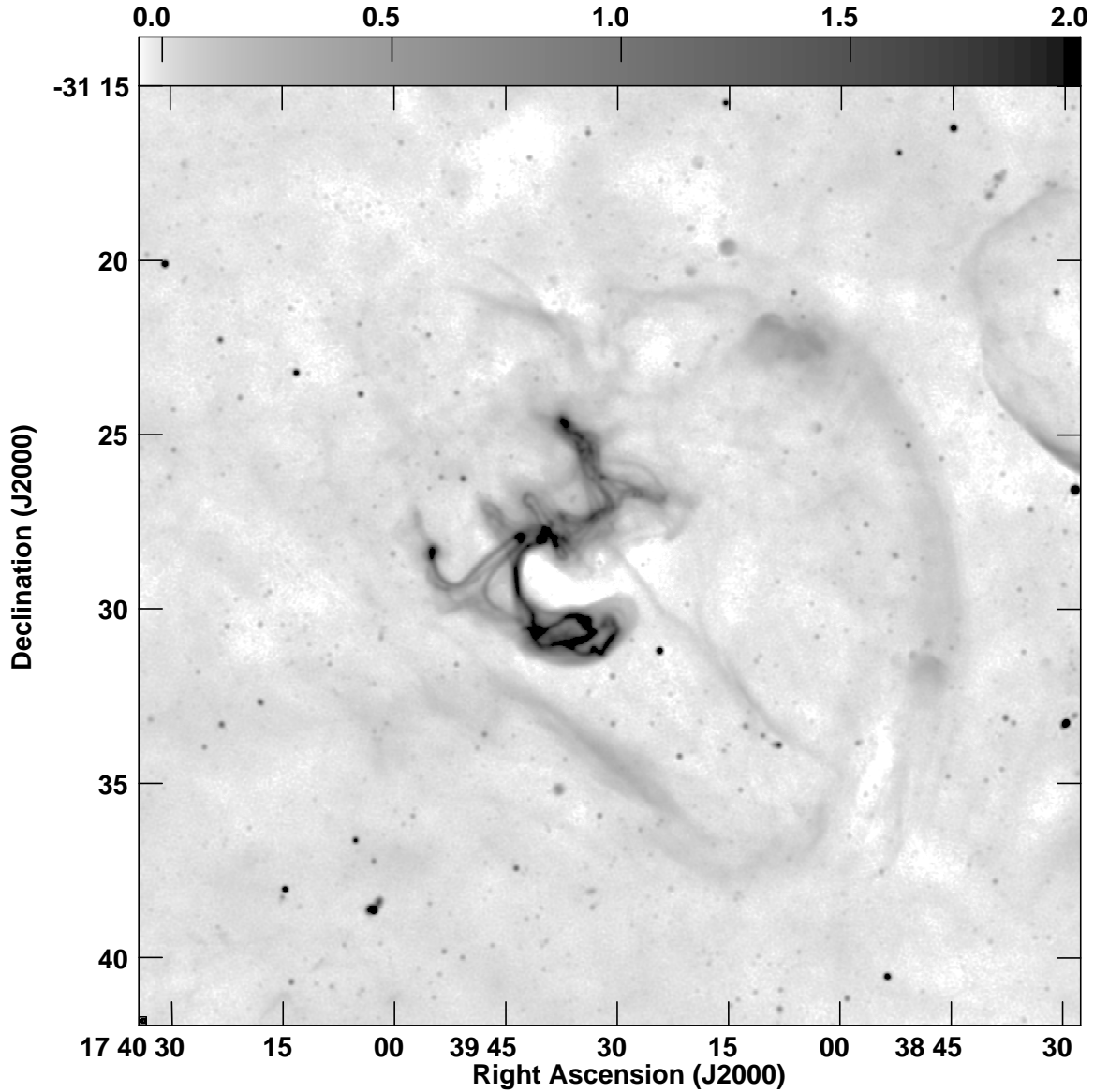}
\caption{Reverse gray-scale of the L-band \chg{(886--1681\,MHz)} Stokes~I image of G357.2$-$0.2
(the Heartworm) in double log stretch with a scale-bar at the top
labeled in mJy\,beam$^{-1}$.  The resolution is shown in the box
at lower left. This rendering optimizes the display of the larger-scale
low brightness emission, including the shell-like heart-shaped
feature spanning $\sim 18\arcmin$ northwards from (R.A., Dec.)
$\approx$ ($17^{\mathrm h}39^{\mathrm m}15^{\mathrm s}$,
$-31\arcdeg38\arcmin$).  The central fine-scale features, considerably
saturated in this view, are best discerned in Figure~\ref{Worm_L}.
} 
\label{Heartworm_L}
\end{figure*}

\begin{figure*}
\plotone{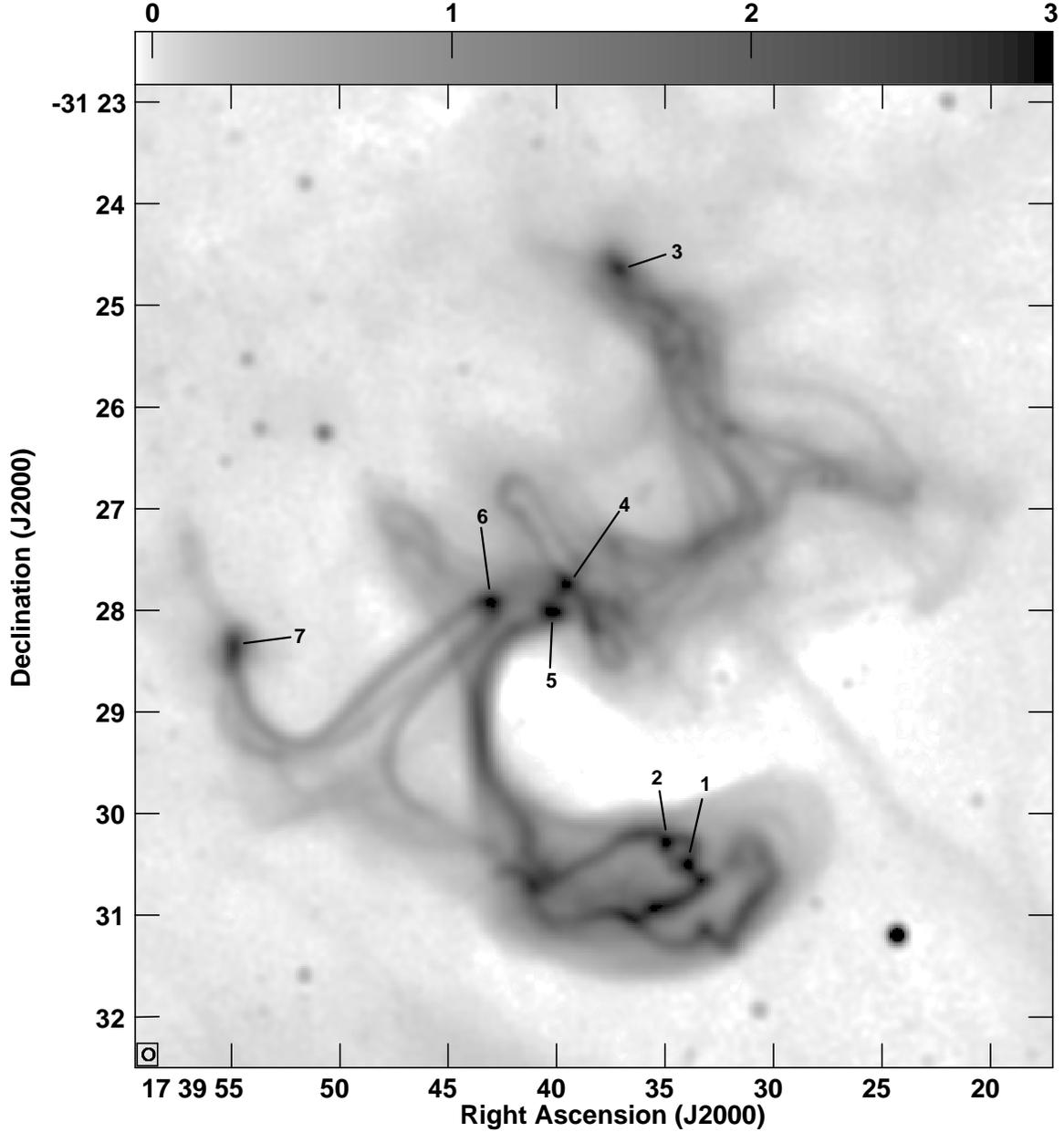}
\caption{Zoom in on Figure~\ref{Heartworm_L}, with a different
contrast (reverse gray-scale in double log stretch with scale-bar
at the top labeled in mJy\,beam$^{-1}$), to highlight the fine-scale
features of G357.2$-$0.2.  The resolution is shown in the box at
lower left. Prominent knots of emission are labeled (see
Table~\ref{tab:knots}).  The bright point source at (R.A., Dec.) =
($17^{\mathrm h}39^{\mathrm m}24^{\mathrm s}$,
$-31\arcdeg31\arcmin12\arcsec$) is the pulsar PSR~B1736$-$31 =
J1739$-$3131.
} 
\label{Worm_L}
\end{figure*}

The MeerKAT L-band total-intensity image of G357.2$-$0.2 is shown
in Figure~\ref{Heartworm_L}.  
The region imaged most prominently contains a complex of filamentary
(worm-like) structures spanning $\sim 8\arcmin$, some of which appear
to terminate in brighter knots; for the first time, some of these
filaments are resolved into striking double tails
(Figure~\ref{Worm_L}).  
There is no overall organization apparent and this fine-scale
structure, at least in projection, is embedded in larger-scale low
brightness emission which contains a large amount of flux density.  

\begin{figure*}
\plotone{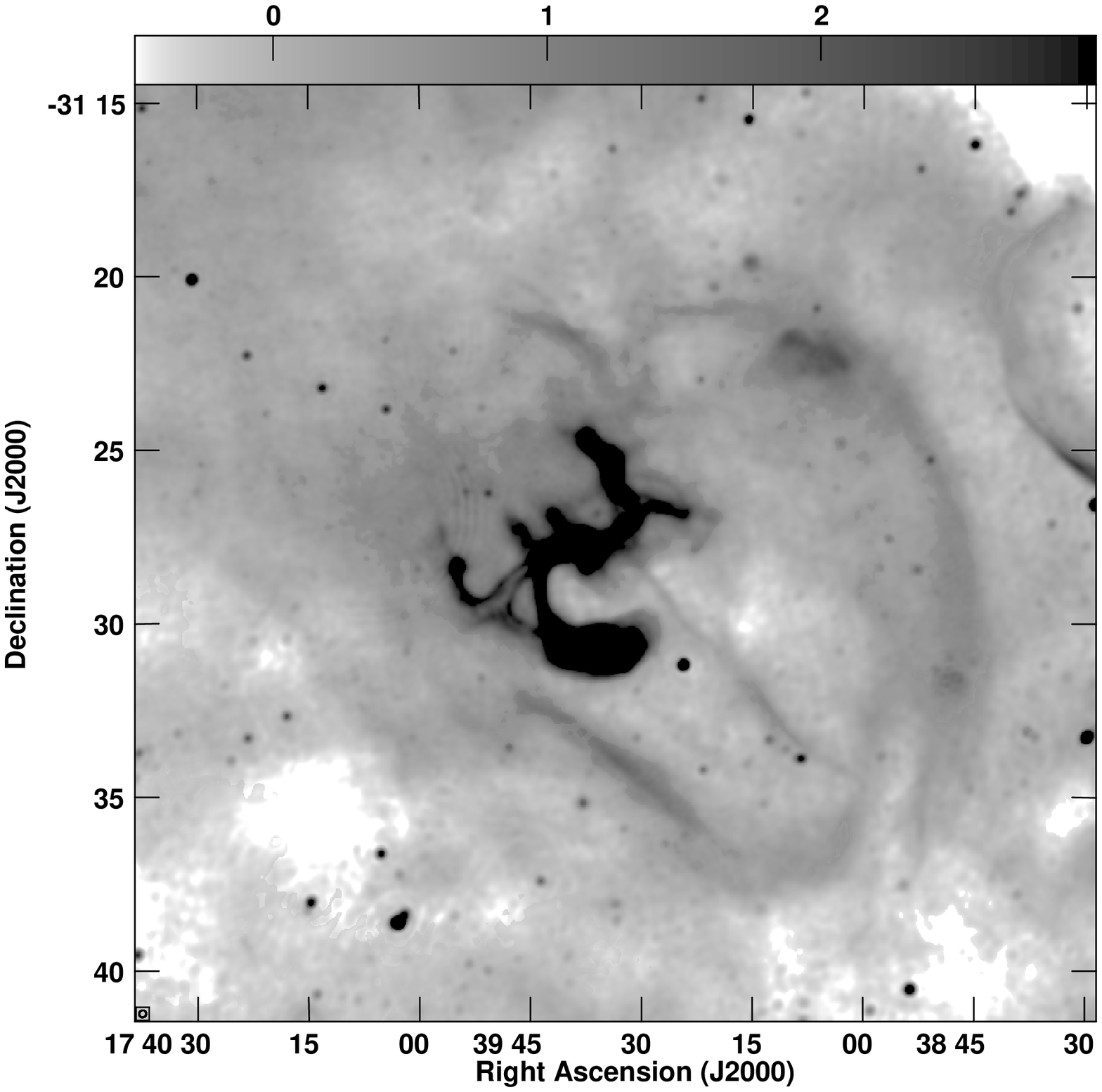}
\caption{\chg{The UHF band (563--1068\,MHz) Heartworm enhanced brightness
sensitivity image in reverse 
gray-scale with double log stretch; a scale-bar is shown at the top labeled in
mJy\,beam$^{-1}$.
The resolution is $12\farcs4\times11\farcs8$ and is shown in the box at
lower left.} 
}
\label{Heartworm_LoRes}
\end{figure*}

\chg{Since the imaging used in Figures~\ref{Heartworm_L} and
\ref{Worm_L} only used the L-band data and explicitly removed the
shorter baselines, the most extended emission is attenuated.
In order to bring out this extended emission, the UHF data were
reimaged with enhanced brightness sensitivity ($\mbox{Robust}=-0.75$) and
including the shorter baselines. 
This is shown in Figure~\ref{Heartworm_LoRes} emphasizing the lower
brightness regions.
}

Some of \chg{the} larger-scale emission appears to be organized in a
partial shell-like heart-shaped feature spanning $\sim 18\arcmin$,
reported here for the first time. 
On the basis of this combined morphology, we have nicknamed these
features the ``Heartworm'' Nebula. 

\subsection{Spectral Index}

\begin{figure}
\centerline{
\includegraphics[width=3.25in]{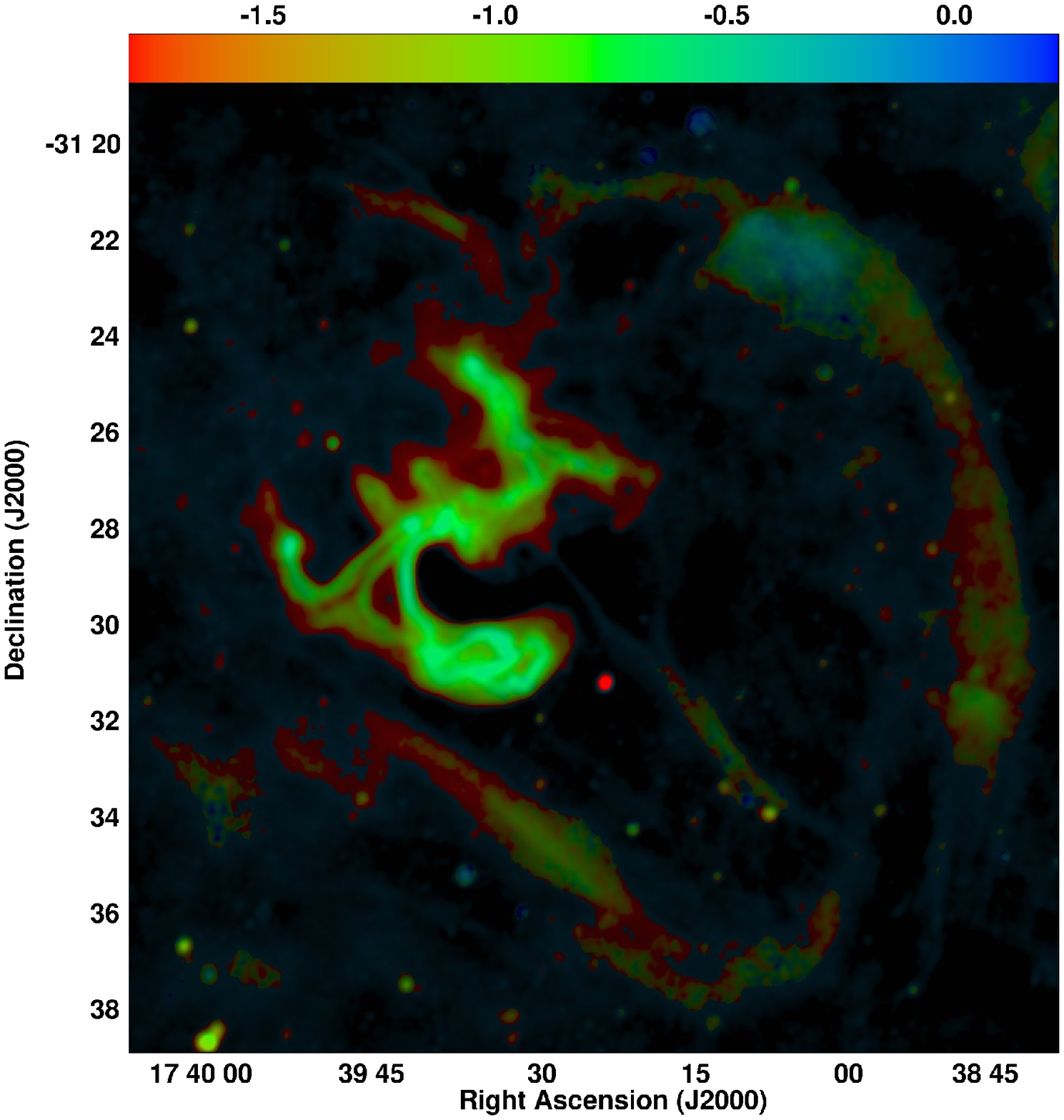}
}
\caption{The spectral index of the Heartworm (G357.2$-$0.2). Intensity
is flux density at 1000\,MHz with square root stretch and color is
spectral index as given by the scale-bar at the top. PSR~B1736$-$31,
with a typical steep pulsar spectrum, corresponds to the prominent
red point. See Figure~\ref{Heartworm_SI_err} for the corresponding
error map.
} 
\label{Heartworm_SI}
\end{figure}

\begin{figure}
\centerline{
\includegraphics[width=3.25in]{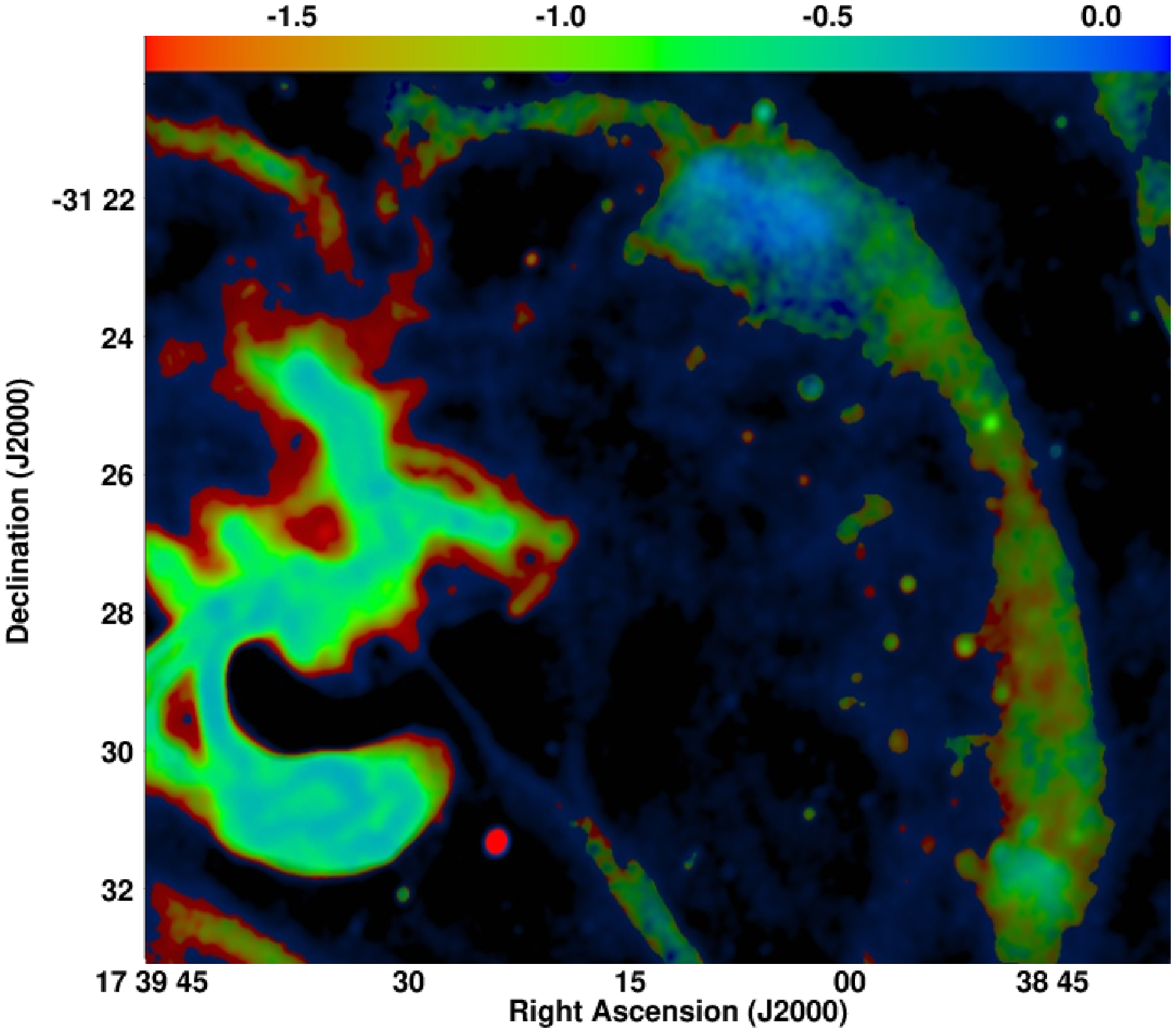}
}
\caption{Like Figure~\ref{Heartworm_SI} but emphasizing the northern
and western rim of the heart.
} 
\label{Heart_SI}
\end{figure}

The individual total-intensity frequency-bin images in the UHF and
L-band images were convolved to a common resolution (that of the
UHF image \chg{described in Section~\ref{StokesIimaging}}) and
interpolated to the grid of the L-band image.  After 
primary beam correction using the frequency-dependent antenna beam
shape of \cite{DEEP2}, a spectrum was fitted in each pixel with the
flux density at 1000\,MHz $S_{\rm 1\,GHz}$ and the spectral index
$\alpha$. The spectral index image is displayed in
Figure~\ref{Heartworm_SI}.  The northern and western rim of the
heart are shown in more detail in Figure~\ref{Heart_SI}.

\begin{figure}
\centerline{
\includegraphics[width=3.25in]{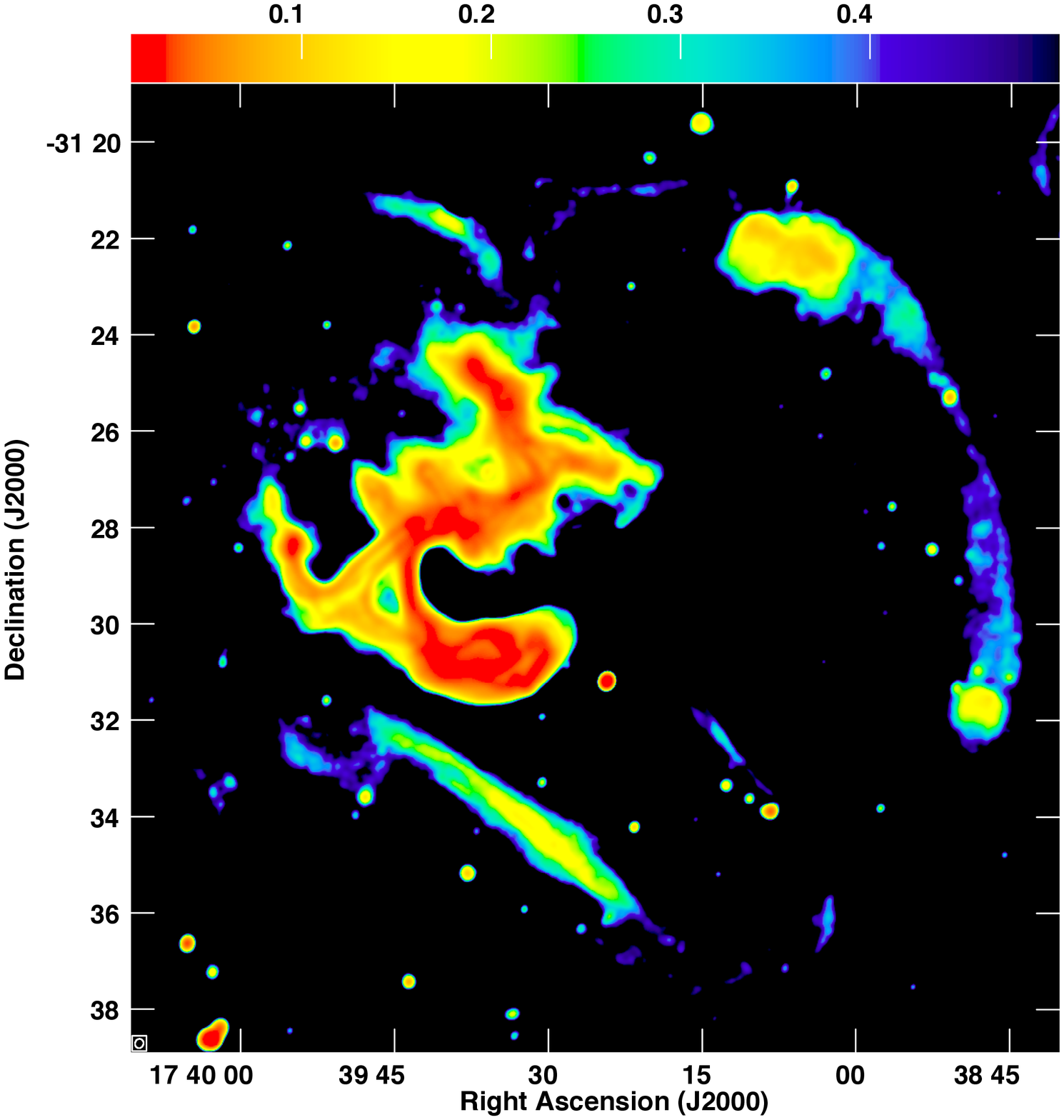}
}
\caption{The error map for the spectral index of the Heartworm
(G357.2$-$0.2) shown in Figure~\ref{Heartworm_SI}.  Color represents
the statistical uncertainty on the spectral index as given by the
scale-bar at the top.  This is based only on RMS noise, and does
not account for systematic errors related to missing flux (see
Section~\ref{StokesIimaging}) or calibration.
} 
\label{Heartworm_SI_err}
\end{figure}

The uncertainty in the spectral index depends on both the signal-to-noise
ratio of a feature across the observed band and any systematics
such as the frequency dependent ``missing'' flux density from
strongly resolved extended emission (see Section~\ref{StokesIimaging}).
The spectral index error image, based only on the statistical
uncertainty, is displayed in Figure~\ref{Heartworm_SI_err}.

\subsection{Polarimetry}

The imaging in Stokes Q and U used 68 $\times$ 1\% fractional
band-pass image planes although many were completely blanked due
to the editing of RFI.  A rotation measure (RM) fit was performed in each pixel by
doing a direct search in Faraday space.  The test Faraday rotation
that gives the highest averaged, unwrapped polarized intensity was
taken as the Faraday rotation at that pixel, the unwrapped polarization
angle extrapolated to zero wavelength was taken as the intrinsic
polarization angle, and the maximum polarized intensity taken as the
polarized intensity in that pixel.  This is essentially taking the
peak of the Faraday synthesis \citep{RMSynthesis}.

\begin{figure}
\centerline{
\includegraphics[height=3.25in]{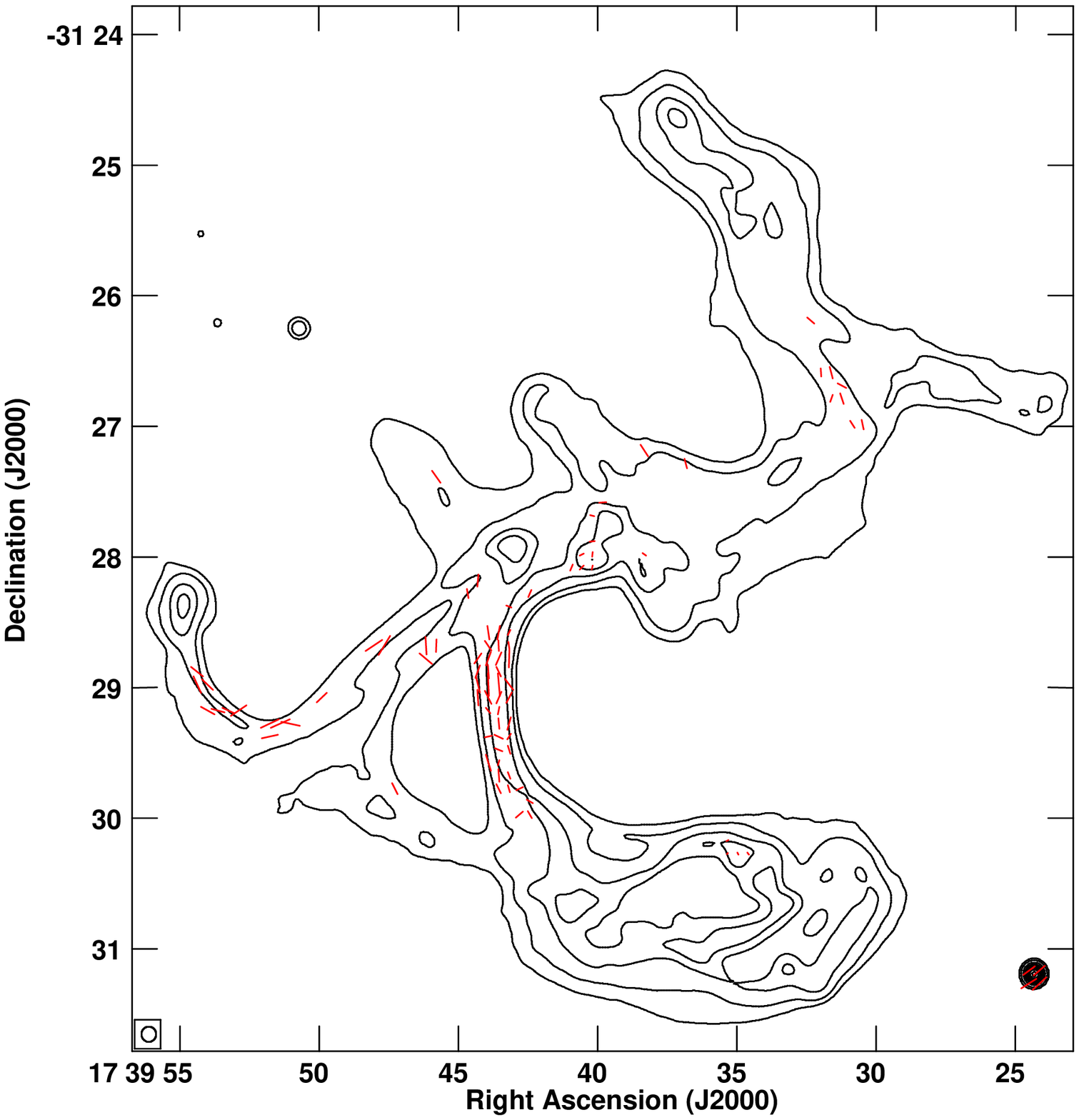}
}
\caption{Total intensity contours of the worm in G357.2$-$0.2, with
superposed red fractional polarization ``B'' vectors from the L-band
data.  Contours are at 2, 4, 8, 12 and 16 $\times$ 0.2\,mJy\,beam$^{-1}$,
and a vector length of $10''$ corresponds to 28\% polarization.
The resolution is shown in the box in the lower left corner.
} 
\label{Heartworm_PolVec}
\end{figure}

\begin{figure}
\centerline{
\includegraphics[width=3.25in]{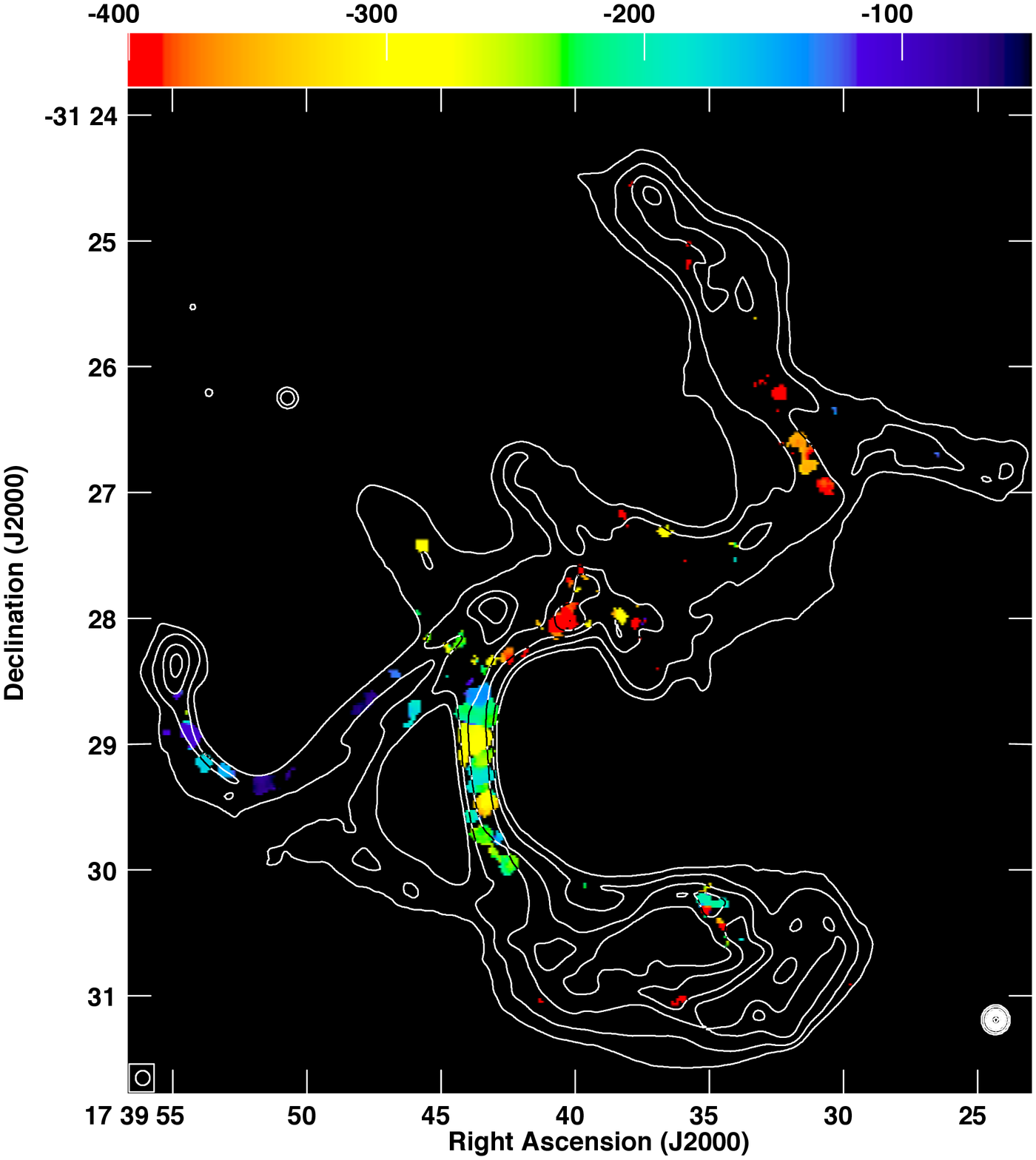}
}
\caption{Total intensity contours of the worm in G357.2$-$0.2, with
superposed RM in color with scale-bar at the top in rad\,m$^{-2}$.
The spotty and highly variable nature of the detected Faraday
rotation suggests that the foreground screen is quite dense and we
are seeing through gaps.  Contours are at 2, 4, 8, 12 and 16 $\times$
0.2\,mJy\,beam$^{-1}$.  The resolution is shown in the box in the
lower left corner.
} 
\label{Heartworm_RM}
\end{figure}

Fractional polarization ``B'' vectors in the worm are shown in
Figure~\ref{Heartworm_PolVec} and the RMs in Figure~\ref{Heartworm_RM}.
Polarization was detected only in limited areas but with moderately
high fractional polarization (20--30\%) and with the magnetic field
largely along the linear features and with large and variable Faraday
rotation.
The rotation measures shown in Figure~\ref{Heartworm_RM} are much less
than the 2000\,rad\,m$^{-2}$ at $\lambda = 6$\,cm found by \cite{Gray96},
supporting the suggestion in the \chg{Figure~\ref{Heartworm_RM}}
caption that at L band and 
UHF we are seeing only through gaps in the dense foreground screen. 

\subsection{X-ray Image} \label{sec:X-ray-image}

\begin{figure}
\centerline{
\includegraphics[width=3.25in]{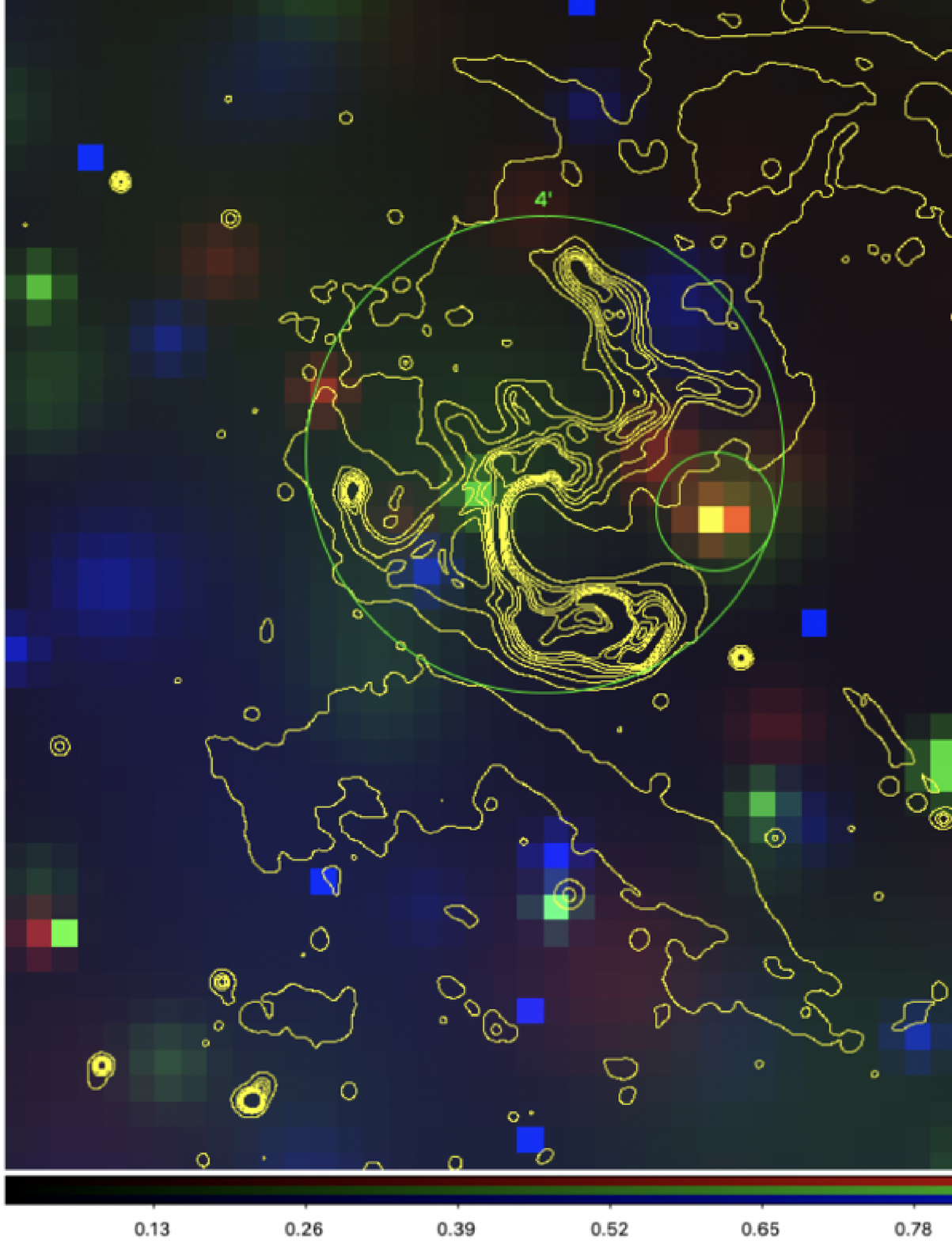}
}
\caption{Three-color image of G357.2$-$0.2 as seen in the eROSITA
all-sky surveys eRASS:4. Photons to produce the image were color
coded according to their energy (red for energies 0.2--0.7\,keV,
green for 0.7--1.2\,keV, blue for 1.2--2.4\,keV).  An adaptive
kernel smoothing algorithm was applied to the images in each energy
band. Radio contour lines (yellow) from the image in
Figure~\ref{Heartworm_L} are overlaid to outline G357.2$-$0.2.  The
green circle with radius 240\arcsec\ encompasses the worm, with a
faint unrelated soft point source located towards the southwest,
indicated by a circle of radius 60\arcsec.
}
\label{fig:eROSITA}
\end{figure}

Figure~\ref{fig:eROSITA} depicts a three-color image of G357.2$-$0.2
which has been coded according to the energy of the detected X-ray
photons. To produce it, we first created images for the three energy
bands 0.2--0.7\,keV, 0.7--1.2\,keV, and 1.2--2.4\,keV, using data
from all seven telescopes.  The spatial binning in these images was
set to $26\arcsec$ to match eROSITA's FoV-averaged FWHM angular
resolution during survey mode.  In order to enhance the visibility
of diffuse emission in the three-color image while leaving point
sources unsmoothed to the greatest possible extent, we applied the
adaptive kernel smoothing algorithm of \cite{2006MNRAS.368...65E}
with a Gaussian smoothing kernel of $1.5\,\sigma$.

As can be seen from Figure~\ref{fig:eROSITA}, no significant diffuse
emission was detected from G357.2$-$0.2 during eRASS:4. There is
some mixture of very faint soft- (red) to medium-band (green)
emission overlapping with the radio contour lines within the large
green circle, but its significance is estimated to be only at the
$\sim 2.5$--3\,$\sigma$ level. Such low level emission is seen at
various other locations in the wider image of all the merged sky
tiles, making it very speculative to associate this faint emission
with G357.2$-$0.2. The small circle in Figure~\ref{fig:eROSITA}
indicates the position of a weak soft point source, which seems
unrelated to the radio features.

\section{Discussion\label{Discussion}}

The H\textsc{i} observations of \cite{Roy2002} indicate that the
worm in G357.2$-$0.2 is at a distance of at least 6\,kpc, possibly
beyond the Galactic center, and likely of Galactic origin. Hereafter
for the purposes of discussion we assume a distance $d = 8.5$\,kpc.
However it is quite unlike any known class of Galactic object, with
the possible exception of PWNe. The worm has a diameter of
$\sim8\farcm3$ which at the assumed distance is equivalent to
$\sim20$\,pc.

\subsection{(Not) Star Formation}

Infrared observations of the Heartworm indicate that the bulk of
the radio features are unlikely to be related to current star
formation.

\begin{figure}
\centerline{
\includegraphics[width=3.25in]{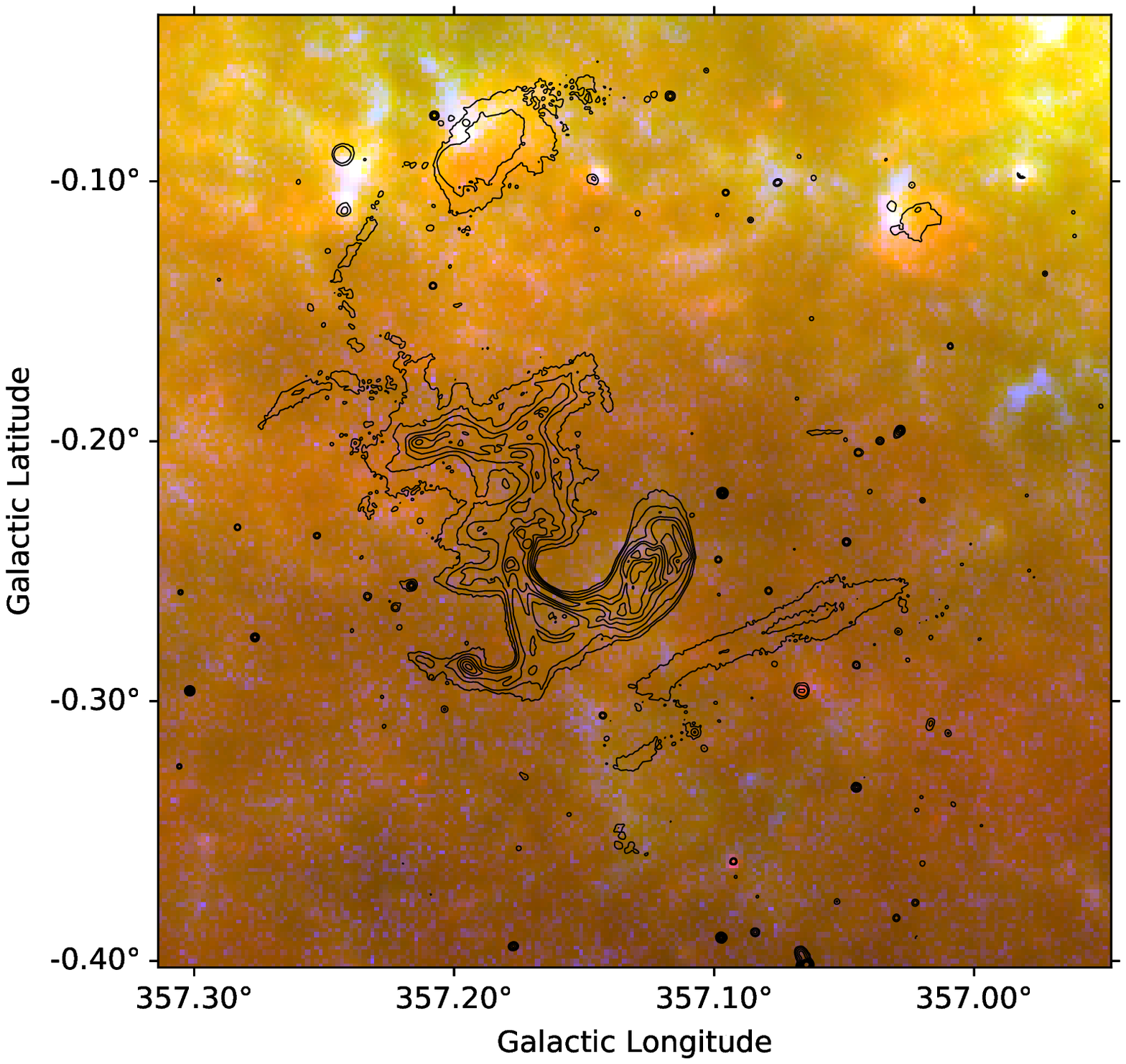}
}
\caption{
Three-color image of the dust emission around the Heartworm Nebula,
in Galactic coordinates and with arbitrary units. Red and green in
the image are respectively coded to PACS 70\,$\mu$m and SPIRE
250\,$\mu$m emission from the \emph{Herschel} Hi-GAL survey
\citep{molinari2010}. Blue is coded to 850\,$\mu$m emission from
the SCUBA-2 Galactic center survey \citep{parsons2018}. Contours
trace the MeerKAT L-band emission from the Heartworm, with levels
chosen using a power-law fitting scheme to emphasise both low-level
and bright emission \citep{thompson2006}. The image shows that there
is little thermal dust emission associated with the worm, although
there is a compact warm dust clump positionally coincident with the
northern end of the heart, indicating a candidate H\textsc{ii}
region, and another such clump and possible H\textsc{ii} region to
its west.
}
\label{3colour}
\end{figure}

There are no extended far-infrared (FIR) features visible near the
worm (Figure~\ref{3colour}) that could be indicative of thermal
dust emission. However, the brightest portion of the heart coincides
with strong FIR emission and may be an H\textsc{ii} region unrelated
to the rest of G357.2$-$0.2 (and hence of unconstrained distance).
This interpretation is supported by the flat radio spectrum of this
region seen in Figure~\ref{Heartworm_SI}. A second smaller clump
of FIR/sub-mm emission may likewise be an unrelated H\textsc{ii}
region (Figure~\ref{3colour}).

\begin{figure}
\centerline{
\includegraphics[width=3.25in]{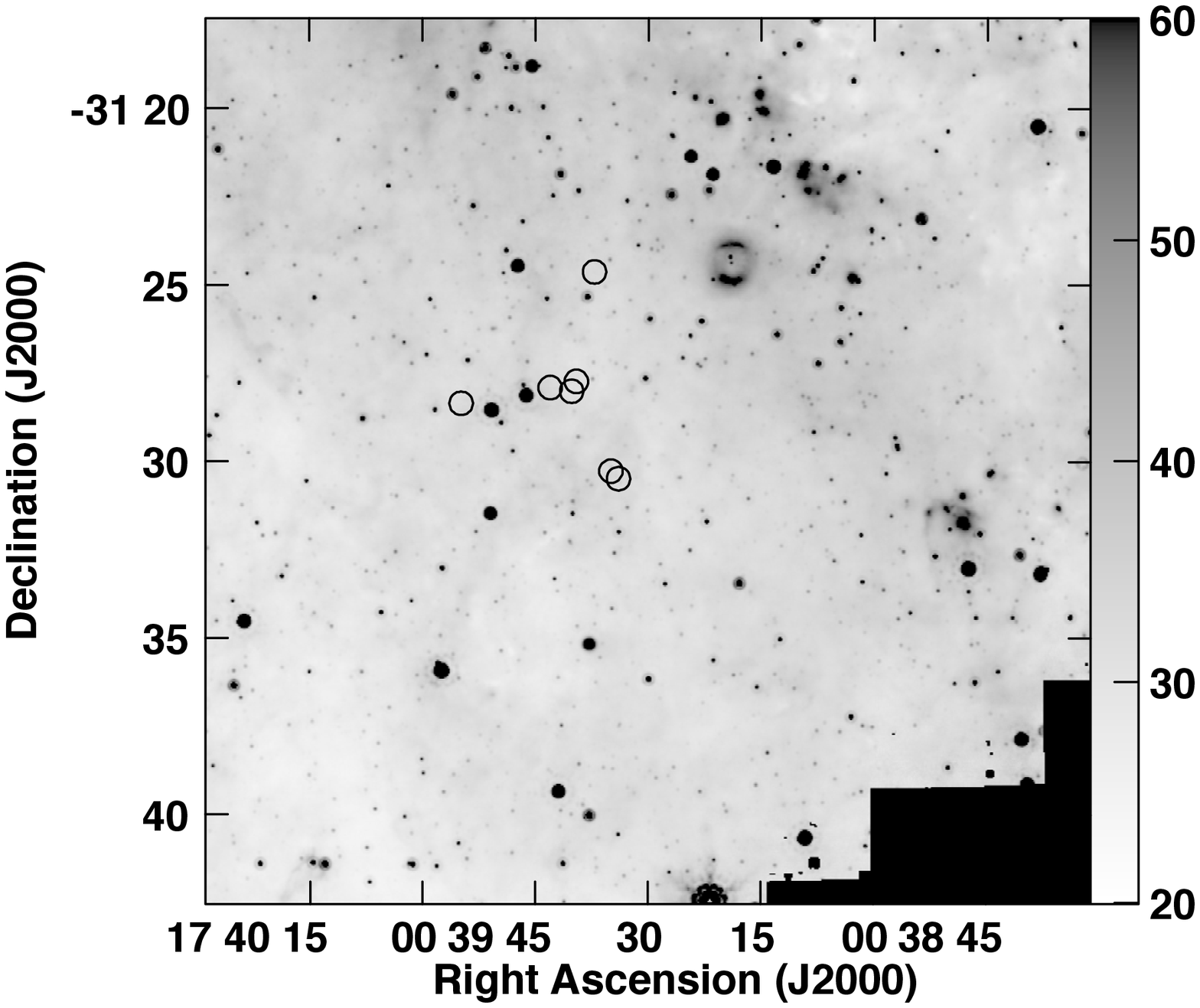}
}
\caption{
The \emph{Spitzer} Enhanced Data Products $24\,\mu$m MIPS image
covering the Heartworm, with circles centered on the knot positions
from Table~\ref{tab:knots} (see also Figure~\ref{Worm_L}). The
intensity scale on the right has units MJy\,sr$^{-1} \approx$
mJy\,beam$^{-1}$.
}
\label{hworm24}
\end{figure}

The strongest argument that the knots in the worm are not H\textsc{ii}
regions is based on the observation that they are fairly strong
radio sources ($S_{\rm 1\,GHz}\sim7$\,mJy according to
Table~\ref{tab:knots}) but are not visible at all ($S_{\rm 24\,\mu
m} \ll 5\,\sigma$) in the deep \emph{Spitzer} Enhanced Data Products
$24\,\mu$m image (Figure~\ref{hworm24}) made with $6\arcsec$ FWHM
resolution. The $24\,\mu$m flux densities of Galactic H\textsc{ii}
regions are typically $30\times$ their 1.4\,GHz flux densities
\citep{Anderson2014} and the $5\,\sigma$ upper limits for sources
smaller than $10\arcsec$ FWHM on the knot positions are $S_{\rm
24\,\mu m} \le 1$\,mJy. Even $A_V = 50$\,mag of extinction would
lower $S_{\rm 24\,\mu m}$ by only a factor of 10 \citep{Anderson2014},
so $< 5\%$ of the knot radio emission is likely to be thermal.

There is also scant indication of correspondence between the compact
radio features in Figure~\ref{Worm_L} and infrared emission at
shorter wavelengths. Knot \#3 is the closest to a near-/mid-infrared
(NIR/MIR) source, with its peak 1\farcs3\ away from a 3.6 and
8\,$\mu$m GLIMPSE-II source \citep{Churchwell2009}.  This source
is also detected in the VVV K$_{\rm s}$ survey but not, as noted
above, in MIPSGAL 24\,$\mu$m.  The \citet{Downes1986} P-statistic
for the possible association of this 8\,$\mu$m 9.259 magnitude
source (the probability of finding a brighter IR source closer to
the radio peak) is $2.3\times10^{-3}$.  Nominally, we might thus
exclude a chance association at the 3\,$\sigma$ level. However this
does not account for MeerKAT astrometric errors, which may contribute
at the $\sim 1\arcsec$ level \citep{Knowles2022,Heywood2022}.  As
for the remaining six radio knots, there are no plausible NIR-MIR
counterparts.

\subsection{The Worm and the Heart}

Both the spectrum and polarized emission suggest that the worm emits
by a nonthermal process, likely synchrotron.  However, the spectrum
of the emission in much of the worm is relatively flat for synchrotron
emission suggesting that the radiating electrons have been recently
accelerated.
\chg{Furthermore}, ionization losses can flatten the spectrum by up to
$\Delta\alpha = +0.5$.

Due to the extended size of the heart, much larger than the $\sim$200\arcsec\
scale filtering in the imaging, much of the emission may be
resolved out.
The rim of this structure survives the filtering of the interferometer
array.  
The spectrum of the bulk of the heart,
at least in the parts of the rim which are well imaged, is relatively
steep (Figures~\ref{Heartworm_SI}--\ref{Heartworm_SI_err}) indicating
an aged relativistic electron population. This excludes the brightest
and flattest-spectrum portion of the heart, which as noted above
may be an unrelated H\textsc{ii} region (see Figure~\ref{3colour}).
Other than positional coincidence, there is no evidence that the
heart and the worm are physically related.

The worm also shares the heart with the pulsar B1736$-$31 (bright
red point in Figure~\ref{Heartworm_SI}) although as already alluded
to in Section~\ref{sec:intro} there is no physical connection between
this pulsar and any of the nearby features. This is further supported
by the RM of the pulsar --- we measure $43.5\pm0.2$\,rad\,m$^{-2}$
\citep[compared to $32\pm8$\,rad\,m$^{-2}$ in][]{Rand94} --- which
is far smaller than that over most of the worm (Figure~\ref{Heartworm_RM}).

\subsection{The Loopy and Knotty Worm}

\begin{figure}
\centerline{
\includegraphics[width=3.25in]{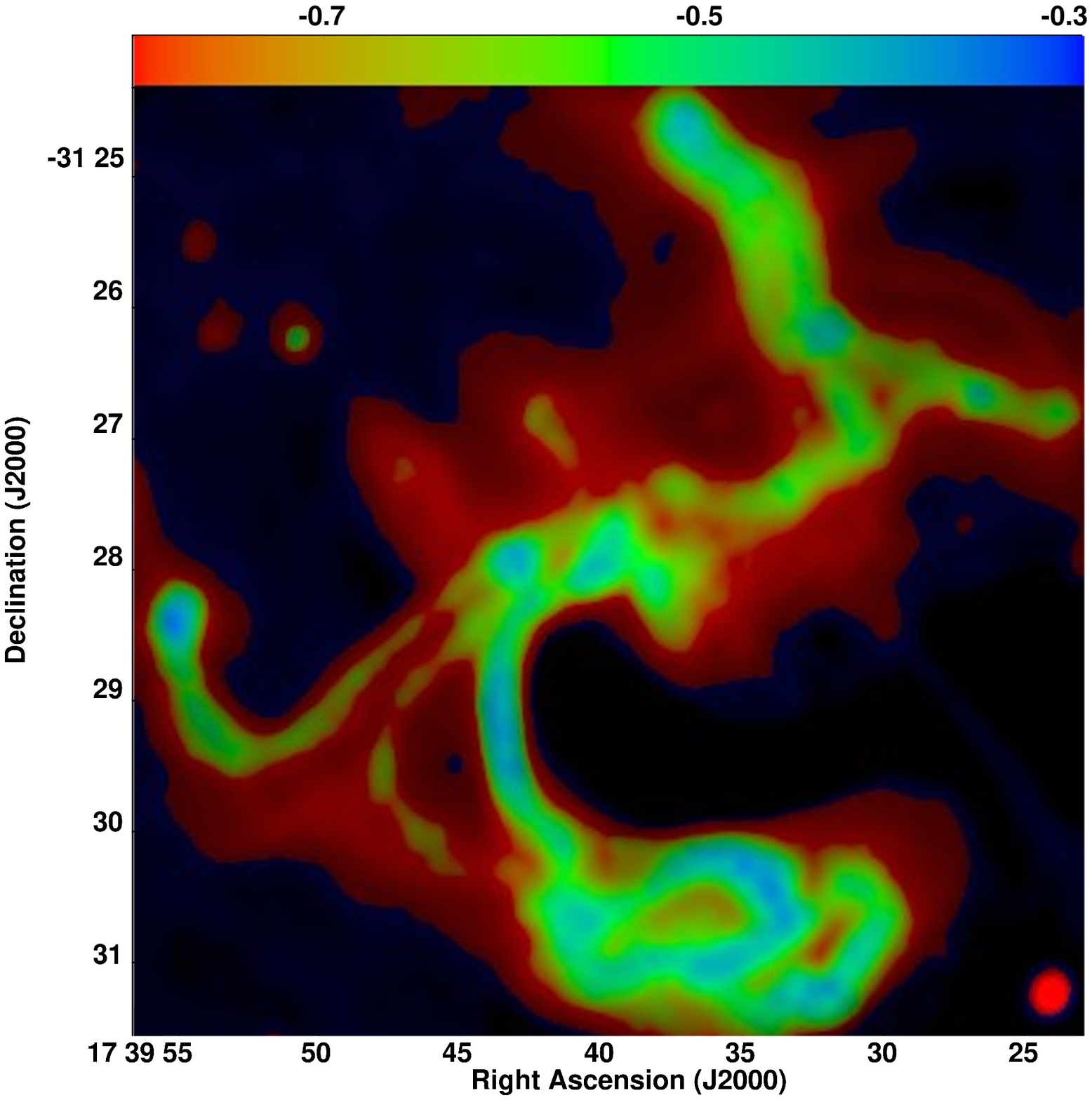}
}
\caption{Like Figure~\ref{Heartworm_SI} but a close up with a tighter
range of spectral index. Note that the region immediately surrounding
the worm appears to have a very steep spectrum ($\alpha \sim -1$),
but this may be affected by the negative bowl due to missing flux
(see Section~\ref{StokesIimaging} and also Figure~\ref{Heartworm_SI_err}).
} 
\label{Heartworm_SI_Close}
\end{figure}

\begin{figure}
\centerline{
\includegraphics[height=3.25in,angle=-90]{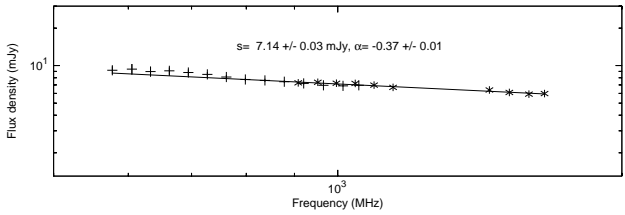}
}
\caption{Spectrum of knot \#6 in Figure~\ref{Worm_L}; see also
Figure~\ref{Heartworm_SI_Close} and Table~\ref{tab:knots}.  UHF
data are displayed as ``+'' and L band as ``*''.  The line is the
fitted spectrum given in the figure, with the flux density provided
for a frequency of 1000\,MHz. Note the match in flux densities
independently determined in the overlapping range $\approx
900$--1050\,MHz.
} 
\label{Heartworm_SI_Point}
\end{figure}

The worm is remarkably complex. Much of its emission seen in
Figure~\ref{Worm_L} consists of filaments.  Many of these are either
paired and connected to a flatter spectrum knot
(Figure~\ref{Heartworm_SI_Close}) or are loops.  Where the polarization
was detectable, the magnetic field appears to be along the filaments
(Figure~\ref{Heartworm_PolVec}) suggesting that they are magnetically
confined structures which have been dragged into their current
configuration, possibly by what is causing the bright knots.  The
flatter spectra near the knots (an example spectrum together with
a least squares fit is given in Figure~\ref{Heartworm_SI_Point})
suggest that these are the locations at which electrons are
accelerated. The identified knots have all very nearly the same
flux densities and nonthermal spectra (Table~\ref{tab:knots}), with
no hint of a break or turnover in the frequency range observed.

\begin{deluxetable}{lrrcc}
\tablewidth{0pt}
\tablecolumns{5}
\tablecaption{Seven knots within the G357.2$-$0.2 nebula}
\tablehead{
        \colhead{\#\tablenotemark{a}} & \colhead{R.A.} & \colhead{Dec.} &
        \colhead{$S_{\rm 1\,GHz}$\tablenotemark{b}} & 
	\colhead{$\alpha$\tablenotemark{b}} \\
	\colhead{} & \colhead{($^{\mathrm h}~^{\mathrm m}~^{\mathrm s}$)} & 
	\colhead{$(\arcdeg\:\arcmin\:\arcsec)$} & \colhead{(mJy)} & \colhead{} 
}
\startdata
1 & 17 39 33.95 & $-31$ 30 29.9 & 7.2 & $-0.37$ \\
2 & 17 39 34.95 & $-31$ 30 16.9 & 7.0 & $-0.36$ \\
3 & 17 39 37.15 & $-31$ 24 38.3 & 6.1 & $-0.39$ \\
4 & 17 39 39.56 & $-31$ 27 44.6 & 7.2 & $-0.43$ \\
5 & 17 39 40.18 & $-31$ 28 01.2 & 7.5 & $-0.39$ \\
6 & 17 39 43.03 & $-31$ 27 55.1 & 7.1 & $-0.37$ \\
7 & 17 39 54.87 & $-31$ 28 21.7 & 5.9 & $-0.33$ \\
\enddata
\tablenotetext{a}{Knots are labeled as in Figure~\ref{Worm_L}.}
\tablenotetext{b}{Flux density values at 1\,GHz and spectral index
$\alpha$ are obtained from pixel-by-pixel fitting over the UHF and
L bands. Uncertainties in $S_{\rm 1\,GHz}$ and $\alpha$ (noise
components only) are $35\,\mu$Jy and 0.02 respectively for each
knot.}
\label{tab:knots}
\end{deluxetable}

There is also a long, relatively straight filament appearing to
connect the center of the worm to the southwestern part of the
heart, at least in projection (see \chg{Figures~\ref{Heartworm_L}--\ref{Heartworm_LoRes}}). 
It is unclear what connection if any this
filament might have to the overall features.

The spotty but high RMs seen in Figure~\ref{Heartworm_RM} and the
strong depolarization reported by \cite{Gray96} indicate that the
emission is behind a relatively dense plasma.  \cite{Gray96} shows
polarized emission at 5\,GHz over most of the worm (Fig.~2) but
reports little polarization at 1.5\,GHz.  The author infers
$\mbox{RM}\sim2000$\,rad\,m$^{-2}$.  This value is substantially
higher than those seen in Figure~\ref{Heartworm_RM}; however, our
resolution is higher than that of \cite{Gray96} at 1.5\,GHz and we
may just be seeing through gaps in an otherwise dense Faraday screen.
Nearby sources, presumed to be background AGNs, have RMs ranging
from $-120$ to +160 rad\,m$^{-2}$ which is outside most of the range
shown in Figure~\ref{Heartworm_RM}, indicating that the bulk of the
Faraday rotation in front of the worm is local to it.

\begin{figure}
\includegraphics[width=3.25in]{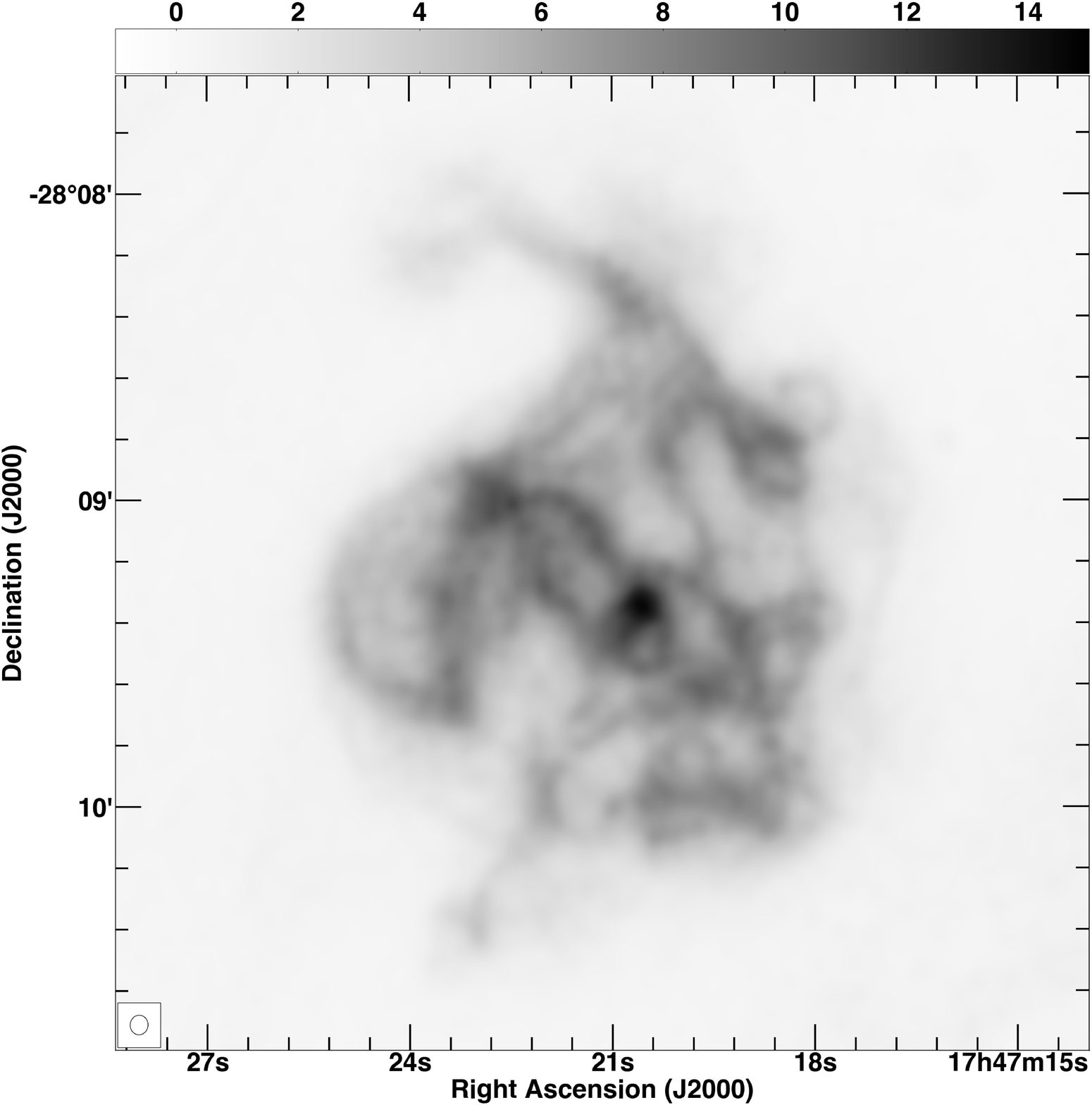}
\caption{
MeerKAT image at 1.28\,GHz showing the PWN at the center of the
SNR~G0.9+0.1. The torus and jet structure inferred from X-ray
observations \citep{gaensler01}, and subsequently reported in radio
imaging by \citet{dubner09}, is revealed here to be a more complex
web of tangled filamentary structures surrounding a prominent central
point-like source \citep[presumably the pulsar discovered
by][]{Camilo2009a}. Compare to the G357.2$-$0.2 worm in
Figure~\ref{Worm_L}.  The angular resolution is 4\arcsec, shown in
the lower left.  The reverse gray-scale is linear with scale-bar at
the top labeled in mJy\,beam$^{-1}$.  Adapted from \cite{Heywood2022}.
} 
\label{G0.9+0.1PWN}
\end{figure}

The filamentary and tangled structure of the worm bears resemblance
to some known PWNe. For instance, the PWN in the composite SNR~G0.9+0.1
(Figure~\ref{G0.9+0.1PWN}) displays a complex web of twisted filaments
(without reported polarization measurements).  By contrast to the
worm, however, no prominent knots of emission are seen in G0.9+0.1.
Conversely, its compact PWN is known to be powered by one of the
most energetic pulsars in the Galaxy \citep{Camilo2009a}, while no
such powering source has been identified for the worm.

\subsection{Pulsar Wind Nebula?}

\subsubsection{The Heartworm as a Composite SNR} \label{sec:g327}

Composite SNRs manifest as a shell (possibly partial and/or distorted)
resulting from the supernova explosion shockwave interacting with
the interstellar medium, together with an interior PWN powered by
a suitably energetic pulsar. The PWNe in middle-aged or older
composite SNRs are often complex in structure due to the fact that
they have been disrupted by the SNR reverse shock (RS). Particularly
in cases for which the shockwave has evolved in a nonuniform medium,
this disruption can result in a complex structure in which the relic
PWN becomes highly distorted \citep{Blondin2001,Kolb2017}, and in
which freshly injected particles and magnetic flux create a new
extended structure near the pulsar. The worm in G357.2$-$0.2, while
unique in some ways, shares several properties of the comparatively
bright PWN in G327.1$-$1.1, which appears to be an example of a
system that has undergone an interaction between the PWN and an
asymmetric RS \citep{Temim2009,Temim2015}.

Australia Telescope Compact Array images of G327.1$-$1.1 taken at
3~cm show diffuse emission from the PWN along with a network of
filamentary structures accompanied by bright knots \citep{Ma2016}.
Accompanying polarization measurements at 6~cm show that the magnetic
field is largely aligned with the filaments.  G327.1$-$1.1 also has
a dense and variable Faraday screen with up to $-600$\,rad\,m$^{-2}$
and an average of $-380$\,rad\,m$^{-2}$ \citep{Ma2016}.  These
features are similar to what is seen in G357.2$-$0.2 in
Figures~\ref{Heartworm_PolVec} and \ref{Heartworm_RM}.

An elongated structure in G327.1$-$1.1 also extends from the putative pulsar ---
identified as an X-ray source with spectral properties consistent
with a neutron star --- back into the relic nebula.  Hydrodynamical
studies show that this appears to be associated with the current
outflow from the pulsar, swept into a tail-like structure by the
RS.  More detailed MHD studies are required to assess whether finer
filamentary structures such as seen in the worm might be formed in
this type of RS/PWN interaction.

If the larger heart structure in G357.2$-$0.2 is considered to be
the shell of an SNR, then assuming a Sedov solution \citep[see,
e.g.,][]{Matthews98} yields an age of about $21\,d_{8.5}^{5/2}
(n_0/E_{51})^{1/2}$\,kyr.  For such a solution, the RS would have
already propagated back to the central regions of the SNR. This is
similar to the age estimate for G327.1$-$1.1 ($\sim 17$\,kyr) at a
distance of 9\,kpc.  The radio spectral index for the entire nebula
in G327.1$-$1.1 is $\alpha \sim -0.3$, typical of PWNe, although
the tail-like structure extending from the pulsar has a steeper
spectrum with $\alpha \sim -0.6$, similar to the filamentary
structures in the worm.

\subsubsection{X-ray Limits}

Pulsars that power appreciable PWNe convert a fraction of their
spin-down luminosity $\dot E$ into nonthermal X-rays.  Here we
investigate whether the limits on X-ray emission obtained from the
eROSITA image presented in Section~\ref{sec:X-ray-image} are
consistent with a PWN interpretation for the worm in G357.2$-$0.2.
In what follows we assume that the absorbing hydrogen column to
G357.2$-$0.2 is $N_{\rm{H}}= 10^{22}$\,cm$^{-2}$. This is the total
average column in the direction of the worm \citep{HI4PI2016}, which
we use in the absence of other constraints.

We calculate limits separately for the presence of a point source,
the putative pulsar powering the PWN, as well as extended emission
from the candidate PWN. In what follows we always report unabsorbed
flux and luminosity limits, i.e., intrinsic to the source after
correction for the assumed absorbing column. All limits are reported
at the $3\,\sigma$ level.

No X-ray point source is detected in eRASS:4 within the bounds of
the presumed PWN, indicated by radio contours inside the large green
circle in Figure~\ref{fig:eROSITA}. We considered two different
emission free spots within this region and obtained a mean cumulative
TM1--TM7 count rate for a putative point source of $<0.059$\,cts\,s$^{-1}$
in the 0.2--8\,keV band.

Pulsars detected in X-rays that power PWNe have power-law spectra
with photon index $\Gamma_{\rm psr}$ in the range 1.0--2.7 \citep[see,
e.g.,][]{Becker09}. Here we assume $\Gamma_{\rm psr} = 1.7$
\citep[e.g., applicable to PSR~J2021+3651 with $\dot E =
3\times10^{36}$\,erg\,s$^{-1}$,][]{Hessels2004}.  For this spectrum,
the above count rate limit yields $f_x(0.2-8\,\mbox{keV}) <
1.3\times10^{-13}$\,erg\,s$^{-1}$\,cm$^{-2}$ for the unabsorbed
energy flux of an undetected point source. For comparison with a
more commonly referenced band, $f_x(0.2-2.4\,\mbox{keV}) <
7.9\times10^{-14}$\,erg\,s$^{-1}$\,cm$^{-2}$. Using the assumed
$d=8.5$\,kpc for G357.2$-$0.2, we estimate that the isotropic X-ray
luminosity of the undetected putative neutron star is $L_{x, \rm
psr} = 4 \pi d^2 f_x < 6.9 \times 10^{32}$\,erg\,s$^{-1}$ within
the 0.2--2.4\,keV band.

The observed nonthermal X-ray efficiency of rotation-powered pulsars
($\eta_{x, \rm psr} \equiv L_{x, \rm psr} / \dot E$) clusters around
$10^{-3}$ in the 0.1--2.4\,keV band
\citep[see][]{1997A&A...326..682B,Becker09}. The above point source
limit therefore nominally implies $\dot{E} < 6.9\times
10^{35}$\,erg\,s$^{-1}$. Given the scatter in the $\eta_{x, \rm
psr}$ relation, and the uncertainties in $N_{\rm H}$ and $d$, this
limit does not exclude the existence of a pulsar of intermediate
$\dot E \sim 10^{36}$\,erg\,s$^{-1}$ powering G357.2$-$0.2 and
beaming towards the Earth. Also, it is always possible that an
unfavorable beaming geometry would preclude direct detection of
nonthermal emission from a pulsar regardless of $\dot E$ and
sensitivity.  However, regardless of geometry a suitably energetic
pulsar should manifest itself via a diffuse PWN.

To constrain extended X-ray emission from G357.2$-$0.2, we derived
the count rate limit within the circle of radius 240\arcsec\ in
Figure~\ref{fig:eROSITA}, which encompasses most of the putative
radio PWN, after subtracting the contribution from the faint
southwestern point source. We obtained a cumulative count rate
$<0.18$\,cts\,s$^{-1}$ in the 0.2--8\,keV band.

PWNe detected in X-rays have power-law spectra with $\Gamma_{\rm
pwn}$ in the range 1.0--2.2 \citep[see, e.g.,][]{2008AIPC..983..171K}.
Here we assume $\Gamma_{\rm pwn} = 2.0$ (e.g., applicable to the
G327.1$-$1.1 PWN discussed in Section~\ref{sec:g327}). For this
spectrum, the above count rate limit gives $f_x(0.2-8\,\mbox{keV})
< 4.1\times10^{-13}$\,erg\,s$^{-1}$\,cm$^{-2}$.  In turn, with
$d=8.5$\,kpc we obtain $L_{x, \rm pwn} = 4 \pi d^2 f_x < 3.6 \times
10^{33}$\,erg\,s$^{-1}$ for the putative PWN in G357.2$-$0.2.

The observed X-ray efficiency of PWNe spans a wide range, with the
bulk within $10^{-5} < \eta_{x, \rm pwn} < 10^{-2}$
\citep{2008AIPC..983..171K}. In any case, there are many instances
of X-ray PWNe powered by pulsars with $\dot E = 10^{36-37}$\,erg\,s$^{-1}$
(e.g., PSR~J2021+3651 and Vela) that have $L_{x, \rm pwn}$ below
our limit for G357.2$-$0.2, and a few such instances powered by
pulsars with even higher $\dot E$ \citep[e.g.,
PSR~J2229+6114,][]{Halpern2001}.

Therefore, the current X-ray limits\footnote{We have also analyzed
\emph{Swift} X-Ray Telescope observations of this region resulting
in the concatenated image available at
\url{https://www.swift.ac.uk/2SXPS/Fields/10000013359}.  No sources
are detected and the limits at the location of G357.2$-$0.2 are 5
times poorer than those from the eROSITA observations.} do not rule
out that G357.2$-$0.2 may be powered by a pulsar of intermediate
$\dot E$, like many that power a variety of PWNe.

For completeness, we also searched the \emph{Fermi}-LAT 4FGL catalog
\citep{Fermi4FGL} for a source coincident with G357.2$-$0.2 but
there are none. This is not constraining: while many energetic
pulsars emit in GeV $\gamma$-rays, their $\dot E/d^2$ flux needs
to be large \citep{Fermi2PC}.

\section{Summary\label{Summary}}

G357.2$-$0.2 consists of two possibly related components, the
``worm'', a series of filaments; and the ``heart'' which is an
extended heart-shaped feature of which we may only see the rim.
H\textsc{i} observations of \cite{Roy2002} show the worm to be of
Galactic origin.  The pulsar B1736$-$31 appears inside the heart
but is a chance positional coincidence.  Part of the rim of the
heart appears to be an unrelated H\textsc{ii} region.

The spectrum and polarization of the emission indicate that the
bulk of the emission from both the worm and the heart is nonthermal
synchrotron.  There is a dense plasma, possibly associated with the
heart, that results in a large Faraday rotation and some depolarization
of the emission from the filaments of the worm.  These appear to
be magnetic structures lit up by particle acceleration in knots
which are associated with the filaments and which appear to be
dragging the magnetic field tubes.  The nature of these knots is
uncertain.

The structure of the worm at least superficially resembles some
PWNe with much of the emission appearing in the form of tangled
filaments.  More sensitive X-ray observations are of particular
interest to further understand the nature of this source.  MeerKAT
observations at S band, with higher angular resolution and less
susceptible to depolarization, may also be instructive. In addition,
detailed hydrodynamical studies could be revealing.  An ultra-deep
radio pulsar search might also be illuminating \citep[see][]{Camilo2009b}.
Nevertheless, if close to the Galactic center, this $\sim20$\,pc
structure would be a very large PWN.  The possibility remains that
this is a more exotic object, perhaps sculpted in part by interaction
with outflows from the Galactic center region.

The radio imaging products presented here are made available with
this article\footnote{\url{https://doi.org/10.48479/q20r-hb79}},
\chg{including Stokes~I (L band, UHF+L band, UHF enhanced surface
brightness sensitivity),
spectral index (UHF+L band), and Stokes Q and U L-band cubes.} Raw
visibility products are available from the MeerKAT data
archive\footnote{\url{https://archive.sarao.ac.za}} under project
code SSV-20200720-SA-01.

\acknowledgments
\chg{We would like to thank the anonymous reviewer for helpful comments.}
The MeerKAT telescope is operated by the South African Radio Astronomy
Observatory which is a facility of the National Research Foundation,
an agency of the Department of Science and Innovation.
The National Radio Astronomy Observatory is a facility of the National
Science Foundation, operated under a cooperative agreement by Associated
Universities, Inc.
MAT acknowledges support from the UK's Science \& Technology Facilities
Council [grant number ST/R000905/1].
eROSITA is the primary instrument aboard SRG, a joint Russian-German
science mission supported by the Russian Space Agency (Roskosmos),
in the interests of the Russian Academy of Sciences represented by
its Space Research Institute (IKI), and the Deutsches Zentrum f\"ur
Luft- und Raumfahrt (DLR). The SRG spacecraft was built by Lavochkin
Association (NPOL) and its subcontractors, and is operated by NPOL
with support from IKI and the Max Planck Institute for Extraterrestrial
Physics (MPE).  The development and construction of the eROSITA
X-ray instrument was led by MPE, with contributions from the Dr.~Karl
Remeis Observatory Bamberg \& ECAP (FAU Erlangen-N\"urnberg), the
University of Hamburg Observatory, the Leibniz Institute for
Astrophysics Potsdam (AIP), and the Institute for Astronomy and
Astrophysics of the University of T\"ubingen, with the support of
DLR and the Max Planck Society.  The Argelander Institute for
Astronomy of the University of Bonn and the Ludwig Maximilians
Universit\"at Munich also participated in the science preparation
for eROSITA. The eROSITA data shown here were processed using the
eSASS/NRTA software system developed by the German eROSITA consortium.

\facilities{MeerKAT, eROSITA}


\software{\emph{Obit} \citep{Obit}}



\clearpage

\bibliography{Heartworm_v2}{}
\bibliographystyle{aasjournal}



\end{document}